\documentclass[aps,pre,superscriptaddress,twocolumn,showpacs]{revtex4}
\usepackage{epsfig}

\newcommand{\Equ}[1]{(\ref{#1})}
\newcommand{\be}{\begin{equation}}
\newcommand{\ee}{\end{equation}}
\newcommand{\bc}{\begin{center}}
\newcommand{\ec}{\end{center}}
\newcommand{\bi}{\begin{itemize}}
\newcommand{\ei}{\end{itemize}}
\newcommand{\ba}{\begin{eqnarray}}
\newcommand{\ea}{\end{eqnarray}}
\newcommand{\ie}{{\it i.e. }}

\newcommand{\ignore}[1]{}

\begin{document}

\title{Co-evolution of dynamical states and interactions in dynamic networks}

\author{Mart\'\i n G. Zimmermann}
\affiliation{Instituto Mediterr\'aneo de Estudios Avanzados
IMEDEA (CSIC-UIB), E-07071 Palma de Mallorca, Spain}
\affiliation{Departamento de F\'{\i}sica, Universidad de Buenos Aires,
Buenos Aires, Argentina}
\affiliation{Santa Fe Institute, 1399 Hyde Park Road, Santa Fe,
New Mexico 87501, USA}
\author{V\'\i ctor M. Egu\'\i luz}
\email{victor@imedea.uib.es}
\affiliation{Instituto Mediterr\'aneo de Estudios Avanzados
IMEDEA (CSIC-UIB), E-07071 Palma de Mallorca, Spain}
\author{Maxi San Miguel}
\affiliation{Instituto Mediterr\'aneo de Estudios Avanzados
IMEDEA (CSIC-UIB), E-07071 Palma de Mallorca, Spain}

\date{\today}

\begin{abstract}

We explore the coupled dynamics of the internal states of a set of
interacting elements and the network of interactions among them.
Interactions are modeled by a spatial game and the network of
interaction links evolves adapting to the outcome of the game. 
As an example we consider a model of cooperation, where the 
adaptation is shown to facilitate the formation of a 
hierarchical interaction network that sustains a highly 
cooperative stationary state. The resulting network has the 
characteristics of a small world network when a mechanism of local 
neighbor selection is introduced in the adaptive network dynamics. 
The highly connected nodes in the hierarchical structure of the 
network play a leading role in the stability of the network. 
Perturbations acting on the state of these special nodes trigger 
global avalanches leading to complete network reorganization. 

\end{abstract}
\pacs{89.75.Hc, 02.50.Le, 87.23.Ge, 89.65.-s}

\maketitle

Recent studies on the structure of social, technological and
biological networks have shown that they share salient features
which situate them far from being completely regular or random
\cite{Watts98,Strogatz01,rev}. Most of the models proposed to
construct these networks are grounded in a graph-theoretical
approach, i.e., algorithmic methods to build graphs formed by
elements (the nodes) and links that evolve according to
pre-specified rules. Despite the progress made, there are still
several open questions \cite{Strogatz01}. An important issue to be
considered among these questions is that networks are dynamical
entities \cite{Skyrms00} that evolve and adapt driven by the
actions of the elements that form a network.

The aim of this paper is to analyze a simple setting of such adaptive and
evolving network, in which there is co-evolution of the state of the elements
in the nodes of the network and the interaction links defining the network.
Interactions among elements are modeled with the aid of game theory
\cite{Weibull96}, frequently applied in social, economic and biological
situations. This mathematical theory models an interaction involving two (or
more) elements, each with two or more 'strategies' or states, such that the
outcome depends on the choices of all the interacting elements. The outcome is
given in the form of a `utility' or payoff given to each element according to
the selected action of the interacting elements. The introduction of spatial
interactions lead to the development of 'spatial games'
\cite{Axelrod84,Nowak92,Abramson01}, where the elements are located in the
nodes of a fixed network of interaction, displaying a rich spatio-temporal
dynamics. We go here beyond these studies by introducing adaptation ({\em
plasticity}) in the coupling between elements, so that the network of
interaction evolves adapting to the outcome of the game. Our results include
new asymptotic steady states, and the emergence of a hierarchical network
structure which governs the global dynamics of the system.

{\em The model.-} We consider a system composed of $N$ elements whose
interactions are specified by a network $\cal N$. The neighborhood of element
$i$ , ${\cal V}_i$, is composed by those elements directly connected to $i$ by
one link, and the size of ${\cal V}_i$ defines its {\em degree}
$k_i$. The state of each element $x_i$ can be $(1,0)$ or $(0,1)$. In each step
(generation), every $i$-th element interacts with all other elements inside its
neighborhood ${\cal V}_i$, and accumulates a payoff
$
\Pi_i = \sum_{j\in {\cal V}_i} x_i {\bf J} x^T_j~,
$
depending on the
chosen states $x_i$ and payoff matrix ${\bf
J}=\left(\begin{array}{cc} \pi_{00} & \pi_{01}\\ \pi_{10} &
\pi_{11}\end{array}\right)$. The $i$-th element compares its own
payoff with all $j\in {\cal V}_i$ and changes its state to the
state of the site with the greatest payoff in $\{i\}\cup{\cal
V}_i$ \cite{Nowak92}.
The {\em plasticity} of the network is introduced here as network
dynamics in which existing links can be severed and replaced by
new ones. 
We make the assumption that whether an interaction link is severed
depends on the joint payoff, \ie, the total payoff by the pair of interacting
elements: the interactions giving the lowest benefit will be removed. 

In the remainder, for the sake of concreteness, we will address the case of the
Prisoner's Dilemma (PD) game, which has been widely used as a model displaying
complex behavior emerging from the competition between cooperative and selfish
behavior \cite{Axelrod84}. In its simplest form, two elements may either choose
to cooperate (C, $x_C = (1,0)$), or defect (D, $x_D=(0,1)$). If both elements
choose C, each gets a payoff $\pi_{00}$; if one defects while the other
cooperates, the former gets payoff $\pi_{10} > \pi_{00}$, while the latter gets
the 'suckers' payoff $\pi_{01} < \pi_{00}$; if both defect, each gets
$\pi_{11}$. Under the standard restrictions $\pi_{10} + \pi_{01}<2\pi_{00}$,
$\pi_{10}>\pi_{00}>\pi_{11}>\pi_{01}$, defection is the best choice in a
one-shot game resulting in a Nash equilibrium where both elements defect. 
Following previous studies  \cite{Nowak92,Abramson01}, we consider a
simplified  version of the game given by the interaction matrix  $\pi_{00}=1$,
$\pi_{10}=b$,$\pi_{11}=\epsilon$, $\pi_{01}=0$, in the limit $\epsilon=0$
\cite{note1}.

\begin{figure}
\epsfig{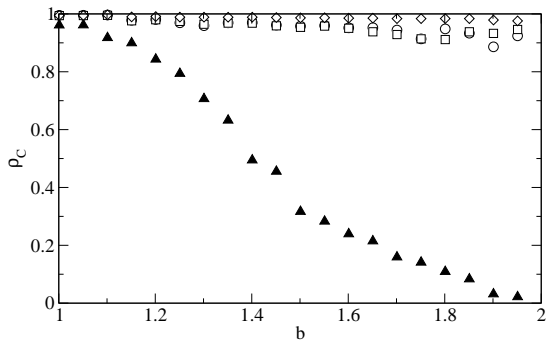}
\caption{
Average fraction of cooperative elements, $\rho_C$, as a function of $b$ and
$p$ in the stationary regime. The defective phase $\rho_C=0$ is not
included in the averages of $\rho_C$. ($p=0$ full triangles, $p=0.01$
circles, $p=0.1$ squares, $p=1$ diamonds)
}
\label{fc}
\end{figure}

In this context the dynamical rule proposed for local neighborhood adaptation,
{\em plasticity}, is defined by analyzing the joint benefit obtained by each of
the possible pairwise interactions: C--C, C--D, and D--D.  Thus, according to
the payoff obtained the worst interaction is clearly observed in a D--D
situation where both elements will be better off by searching a new partner. 
Given this simplistic analysis, taking into account that we are considering
undirected links, and assuming that the probability to rewire a C--D
interaction is much smaller than to rewire a D--D interaction, our
implementation of plasticity will allow Defectors to exchange
(probabilistically) a D-neighbor by another randomly chosen element.

Thus, the game is divided in three stages. (i) Each element $i$ plays the PD
game with the same current state with all its neighbors, and collects an
aggregate payoff $\Pi_i$. (ii) Each element $i$ updates its current state by
comparing its payoff with its neighbors and {\em imitates} the state of the
wealthiest element. An element is said {\em satisfied} if its own payoff is the
highest among its neighbors (otherwise it is unsatisfied). (iii) Unsatisfied
D-elements which imitate a Defector, replace this link with probability $p$ by
a new one pointing to a randomly chosen element.

The plasticity parameter $p$ leads to a time
evolution of the local connectivity of the network leaving
the average degree $\langle k_i \rangle$
constant. The parameter $p$ sets a timescale for the evolution of
the network with respect to the state update. In general we expect
$p \ll 1$, so that the state update evolves in a much faster
timescale than the network evolution, while $p=1$ represents the limit
of simultaneous update of interactions and states.


We have characterized numerically the model using $N=10,000$ elements, averaged
over $100$ different random initial conditions, with an initial population of
$0.6N$ Cooperators randomly distributed in the network \cite{note1}. The initial network is
generated by randomly distributing $N\langle k_i \rangle/2$ links. A prototype
value of $\langle k_i \rangle=8$  was chosen in order to secure an initial
large connected component. The game is played {\em synchronously}, \ie,
elements decide their state in advance and they all play at the same time.

\begin{figure}
\vskip -1.cm
\centerline{\epsfig{file=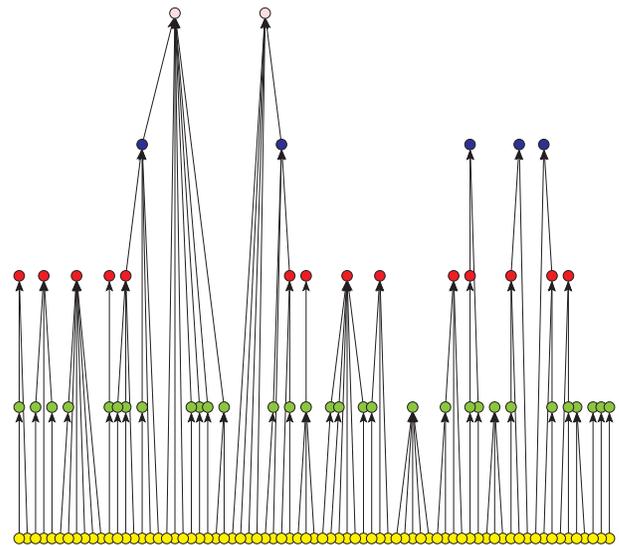,width=0.5\textwidth,angle=-180}}
\caption{\label{struct} Partial view of a sample imitation network in a steady
state. Elements on a lower layer imitate the state of elements in an upper
layer.}
\end{figure}

{\em Stationary states.-}
To characterize the macroscopic behavior of the system we introduce the
fraction of cooperators at a given time, $\rho_C (t)$.
We define the order parameter $\rho_C$ as
the average over realizations of the stationary cooperators density.
In the case of random mixing, \ie in the absence of an interaction
network, population dynamics gives \cite{Sigmund98}
$
\dot \rho_C = \rho_C^2 (1- \rho_C) (1 - b)~.
$
Thus, for $b>1$ the only stable solution corresponds to a fully
defective population.
For fixed networks ($p=0$), a typical time evolution shows in general
that the order parameter fluctuates around an average value that
decreases as the incentive to defect $b$  increases (Fig.~\ref{fc}). At $b
\simeq 2$ the defectors  dominate the network \cite{Lindgren94}. For fixed
networks, the  precise value for this transition has been studied in detail 
\cite{Nowak92,Lindgren94,Schweitzer02}. In contrast to random  mixing, {\it
context preservation} (fixed interactions) {\it  sustains partial cooperation}
\cite{Cohen99}. 

This picture changes when the elements turn on their plasticity behavior
($p>0$) [see Fig.~\ref{fc}]. Extensive numerical simulations show that the
system either reaches a stationary configuration with $\rho_C>0$ (where the
states and the network do not change in time), or an absorbing state with all
elements being Defectors $\rho_C=0$. The {\em cooperative phase} --the
stationary states with a large value of $\rho_C$--  is formed by a set of solutions,
corresponding to different network configurations and distribution of
cooperators. In Fig.~\ref{fc} we characterize these states showing that
$\rho_C>0.8$, a value always much larger than in the non-adaptive case. 
Slight variations exist for different $\langle k \rangle \ge 4$.
The crucial difference is the disappearance of the behavior observed in the
case $p = 0$ in which, increasing $b$, the large majority of the realizations
reaches a configuration with a very low fraction of cooperators. The plasticity
parameter $p$ changes the time it takes to reach the stationary state: smaller
$p$ produce longer transients.  

\begin{figure}
\epsfig{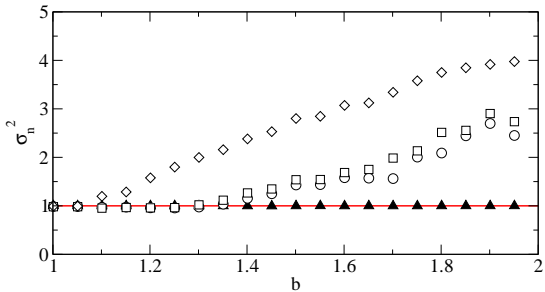}
\caption{Normalized variance
$\sigma_n^2 = (\langle k_i^2 \rangle -\langle
k_i \rangle^2)/ \langle k_i \rangle$ as a function of $b$. 
The solid line ($\sigma_n^2=1$) correspond to the fixed random network with a
Poisson distribution of degree. Parameter values as in Fig.~\ref{fc}.}
\label{hist}
\end{figure}

{\em Network structure.-}
In order to understand how such a highly cooperative structure can
be sustained, we analyze the implications of the proposed dynamical
rules in the network structure.
Consider that element $i$ updates its state imitating
the state of element $j$; we
define the correspondence $l:\cal N\rightarrow\cal N$ such that
$l(i)=j$. Focusing only on those links, we identify the {\em imitation
network} as the sub-network composed of directed links $i\rightarrow
l(i)$ (Fig.~\ref{struct}). A {\em necessary and sufficient} condition for a stationary state
($\rho_C>0$, $p \ne 0$) is (a) there are no links between two
Defectors, and (b) each C-neighbor $i$ of a Defector $\gamma$, satisfies
the payoff relation:
\be
\Pi_j > \Pi_\gamma > \Pi_i, \qquad j=l(i)
\label{payoff}
\ee
In other words, in a stationary state {\em all} defectors become satisfied
interacting only with cooperators, while cooperators can be unsatisfied while
imitating from other cooperators.
These steady state conditions naturally imply
that the element with largest payoff in a stationary configuration is a {\em
satisfied} cooperator. In Fig.~\ref{struct} we show a partial view (the nodes
in the lowest level are not shown) of an imitation network, where the nodes in
a layer imitate those elements in an upper layer indicated by the directed
edges. At the top of the figure lie the nodes whose action is imitated by a
chain of Cooperators.

A first characterization of how the structure of the cooperative stationary network 
configurations changes as a function of $b$ and $p$ is obtained 
by measuring the normalized
degree variance $\sigma_n^2 = (\langle k_i^2 \rangle -\langle k_i 
\rangle^2)/ \langle k_i \rangle$ (Fig.~\ref{hist}). We find 
that the degree distribution departs significantly from the 
initial Poisson distribution ($\sigma_n^2=1$) only for large 
values of the plasticity parameter $p$. 
For increasing $b$ the tail of the degree distribution expands and approaches an
exponential form, indicating some elements become more connected
than others (hubs).

\begin{figure}
\centerline{\epsfig{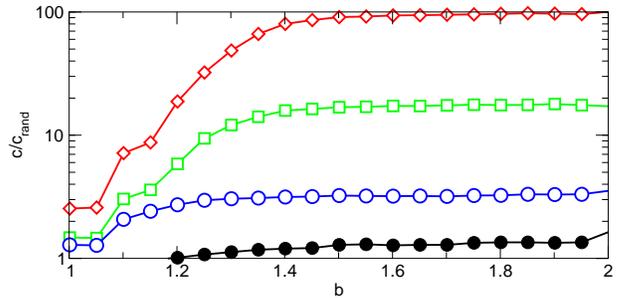}}
\caption{
Normalized clustering coefficient $c/c_{rand}$ as a function of $b$. For $p=0$ we recover the
random value. Open symbols for q=0.01 (diamonds $p=0.01$; squares $p=0.1$;
circles $p=1$); filled circles correspond to $p=1$ and $q=0$.}
\label{C}
\end{figure}

We now address the question of whether the structure generated in our dynamical
model has the characteristics of a {\em small world} network \cite{Watts98}.
The clustering coefficient $c$ measures the fraction of neighbors of a node
that are connected among them, averaged over all the nodes in the network. In
our simulations we find (Fig.~\ref{C}) that the clustering coefficient
increases very mildly with respect to a fixed random network $c_{rand} =
\langle k_i \rangle/N$ \cite{Bornholdt02}. Thus, even though the average path
length is similar to a random network, in order to account for the high
clustering we need to introduce ``local'' neighbor selection \cite{Jin01}. 
This mechanism is easily implemented introducing a parameter $q$ that modifies 
[step (iii)], so that  with probability $q$ the new neighbor is selected among 
the  neighbors of the neighbors; otherwise with probability $1-q$ the  random 
neighbor is chosen.
We find that, while
most of our results previously discussed are qualitatively independent of
the value of $q$, the clustering coefficient reaches a very large value
even for a small value of $q$. For instance, a slightly 1\% ($q=0.01$) of
local neighbor selection is enough to increase $c$ a hundred times, being
the clustering largest for a slow evolution of the network ($p\ll 1$). In
addition, the clustering coefficient decreases slightly with system size,
an indication of a decay slower than the $N^{-1}$ decay expected for
random graphs. All together our results indicate that local
neighbor selection is needed in order to generate a small world network.

It is worth noting that an evolutionary model based on the PD game with a more
complex strategy representation also shows, in the absence of local neighbor
selection, that the increase of the clustering coefficient can be related to
the change of the degree distribution \cite{Bornholdt02}.
In contrast with Ref.~\cite{Bornholdt02}, we don't observe a power law degree
distribution.

\begin{figure}
\centerline{\epsfig{file=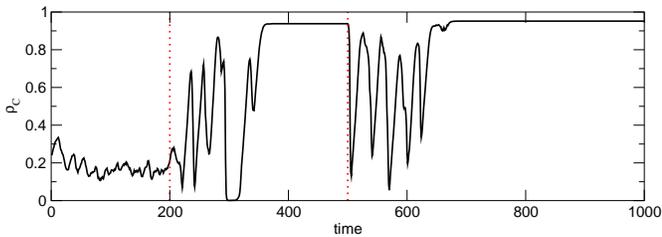,width=0.4\textwidth,angle=-90}}
\vskip -4cm \caption{Time evolution of $\rho_C$. The evolution
starts in a fixed random network ($p=0$ up to time $t=200$ when
network dynamics is switched on, so that $p=1$ for $t>200$. At
time $t=500$ the state of the node with largest payoff is forced 
from C to D. Parameter $b=1.7$.} \label{tlider}
\end{figure}

{\em Dynamics: global avalanches and network stability.-} 
The hierarchical structure of the network is of fundamental
importance to the dynamics on the system. A closer look at the the 
evolution towards a stationary state indicates the presence of avalanches
(Fig.~\ref{tlider}).
The transient dynamics is
characterized in general by large oscillations in $\rho_C(t)$. When the
payoff of any D element increases above the upper limit of
Eq.~\Equ{payoff}, an avalanche towards defection is triggered. 
This D element will be imitated by C neighbors, and each will
initiate an avalanche of replication of D state through all those
elements connected by the imitation network. During the
avalanche, recovery of cooperation is possible through 
those satisfied Cooperators, which re-build the hierarchical 
topology \cite{Zimmermann01}.

The description in terms of the imitation network also indicates the
vulnerability of the structure to stochastic fluctuations. Figure
\ref{tlider} illustrates the sensitivity of the stationary network
structure to perturbations acting on the highly connected nodes, which
reflects their key role in sustaining cooperation. At time $t=500$, the
most connected node is externally forced to change state from C
to D, triggering an avalanche. Notice the large oscillations in
$\rho_C$, reproducing the transient dynamics in which the system
searches for a new stationary globally cooperative structure.

{\em Conclusion.-} 
We have addressed the general question of network
formation from the perspective of co-evolution between the dynamics of
the elements' state and the interactions network.
Our model of cooperation with network
plasticity leads to hierarchical topologies \cite{hier}, the emergence of
global cascades \cite{Jain98,Watts02} and vulnerability to attacks acting
on specific targets \cite{attack}. The hierarchical interaction network
is reached as a stationary network starting form a random network of
interactions. The network appears structured from a few highly connected
elements easily identified through the imitation network. Such network
has the characteristics of a small world when a mechanism of local
neighbor selection is introduced in the adaptive dynamics of the network.
The hierarchical structure supports a stationary highly cooperative state
for general situations in which for a fixed network the system would not
settle in a stationary state and in which the cooperation level would be
much smaller. The stability of the network is very sensitive to changes
in state of the few highly connected nodes: 
external perturbations acting on these nodes, trigger global avalanches
leading to transient dynamics in which the network completely reorganizes
itself searching for a new highly cooperative stationary state.
Future work should explore the robustness of these results in slightly different
settings. For instance we have checked that the same qualitative results are
obtained with asynchronous update regarding Figure~1, and that adding
continuous noise weakens the cooperative phase by  the spontaneous occurrence of
avalanches. Work along these lines is in progress.

We acknowledge financial support from MCyT (Spain) and FEDER (EU)
through Projects CONOCE and BFM2002-04474-C02-01.



\begin{thebibliography}{}

\bibitem{Watts98} D.J. Watts, S.H. Strogatz, Nature {\bf 393}, 440
(1998).

\bibitem{Strogatz01} S.H. Strogatz, Nature {\bf 410}, 268 (2001).

\bibitem{rev} R. Albert A.-L. Barab\'asi, Rev. Mod. Phys. {\bf 74}, 47
(2002); S.N. Dorogovtsev, J.F.F. Mendes, Adv. Phys. {\bf 51}, 1079
(2002); M.E.J. Newman, SIAM Review {\bf 45}, 167-256 (2003).

\bibitem{Skyrms00} B. Skyrms, R. Pemantle, Proc. Natl. Acad. Sci. USA
{\bf 97} 9340 (2000).

\bibitem{Weibull96} J. Weibull, {\sl Evolutionary Game Theory} (MIT
University Press, 1996).

\bibitem{Axelrod84} R. Axelrod, {\sl The Evolution of Cooperation},
Basic Books, New York (1984).

\bibitem{Nowak92} M.A. Nowak, R.M. May, Nature {\bf 359}, 826 (1992);
M.A. Nowak, R.M. May, Int. J. Bif. Chaos {\bf 3}, 35 (1993); M.A.
Nowak, S. Bonhoeffer, R.M. May, Int. J. Bif. Chaos {\bf 4}, 33
(1994).

\bibitem{Abramson01} G. Szabo, C. Toke, Phys. Rev. E {\bf 58}, 69 (1998);
G. Abramson, M. Kuperman, Phys. Rev E {\bf 63}, 030901
(2001); H. Ebel, S. Bornholdt, Phys. Rev. E {\bf 66}, 056118 (2002); B.J. Kim,
et al, Phys. Rev. E {\bf 66}, 021907 (2002). 

\bibitem{note1}
We checked as in Ref.~\cite{Lindgren94} that considering 
$0.1>\epsilon\ge 0$, no noticeble difference is found in this game.

\bibitem{note2} Several intermediate values of the initial density of
Cooperators show that general dynamical properties are not changed. However,
decreasing the initial density of cooperator increases the chances to get
trapped in the absorbing state $\rho_C=0$.

\bibitem{Sigmund98} J. Hofbauer, K. Sigmund, {\sl Evolutionary games
and population dynamics}, Cambridge University Press (1998).

\bibitem{Lindgren94}
K. Lindgren, M.G. Nordahl, Physica D {\bf 75}, 292 (1994).

\bibitem{Schweitzer02} F. Schweitzer, L. Behera, H. Muehlenbein,
Adv. in Complex Systems {\bf 5}, 269 (2002).

\bibitem{Cohen99} M. Cohen, R. Riolo, R. Axelrod, '{\sl The emergence
of social organization in the Prisoner's Dilemma: how
context-preservation and other factors promote cooperation}',
Santa Fe Institute Working Paper 99-01-002 (1999).

\bibitem{Bornholdt02}
H. Ebel, S. Bornholdt, {\tt cond-mat/0211666}.

\bibitem{Jin01}
E.M. Jin, M. Girvan, M.E.J. Newman, Phys. Rev. E {\bf 64}, 046132 (2001).

\bibitem{Zimmermann01}
M.G. Zimmermann, V.M. Egu\'{\i}luz, M. San Miguel,
in {\sl Economics with Heterogeneous Interacting Agents},
A. Kirman, J.-B. Zimmermann eds. (Springer, Berlin, 2001)

\bibitem{hier} E. Ravasz, A.-L. Barabasi, Phys. Rev. E {\bf 67}, 026112 (2003).

\bibitem{Jain98} S. Jain, S. Krishna, Phys. Rev. Lett. {\bf 81},
5684-5687 (1998). S. Jain, S. Krishna, Proc. Natl. Acad. Sci. 2001 {\bf 98},
543 (2001).

\bibitem{Watts02} D.J. Watts, Proc. Natl. Acad. Sci. USA, {\bf 99}, 5766 (2002).

\bibitem{attack} R. Albert, H. Jeong, A.-L. Barabáasi, Nature {\bf 406},
378 (2000); R. Cohen, K. Erez, D. ben-Avraham, and S. Havlin, Phys.
Rev. Lett. 85, 4626 (2000); 86, 3682 (2001).

\end{thebibliography}
\end{document}